\documentclass[aps,prd,12pt,nofootinbib]{revtex4-2}

\usepackage[bookmarksnumbered, pdfpagelabels=true, plainpages=false, colorlinks=true, linkcolor=blue, citecolor=blue, urlcolor=blue]{hyperref}
\usepackage{xcolor}
\usepackage{graphicx}
\usepackage{animate}
\usepackage{bm}
\usepackage{subfloat}
\usepackage{float}
\usepackage{amssymb}
\usepackage{amsmath}
\usepackage{caption}

\newcommand{\be}{\begin{equation}}
\newcommand{\ee}{\end{equation}}
\newcommand{\ba}{\begin{eqnarray}}
\newcommand{\ea}{\end{eqnarray}}
\newcommand{\bd}{\begin{displaymath}}
\newcommand{\ed}{\end{displaymath}}

\def\thalf{{\textstyle{\frac{1}{2}}}}

\def\oneth{{\textstyle{\frac{1}{3}}}}

\def\oneqt{{\textstyle{\frac{1}{4}}}}

\begin{document}

\title{Causally connected regions in relativistic heavy ion collisions}

\author{Olivia Chabowski}
\email{chabo026@umn.edu}
\affiliation{School of Physics \& Astronomy, University of Minnesota, Minneapolis, MN 55455, USA}

\author{Joseph I. Kapusta}
\email{kapusta@umn.edu}
\affiliation{School of Physics \& Astronomy, University of Minnesota, Minneapolis, MN 55455, USA}

\begin{abstract}

Quantifying causal connections within fireballs resulting from high energy nucleus-nucleus collisions is pertinent to assessing the viability of several proposed mechanisms that directly influence particle production and correlations. Fireball causal connections have previously been studied in the context of 1+1 dimensional Bjorken flow. We expand into 3+1 dimensions using Gubser flow, which includes transverse expansion. Our findings suggest that the volume of these causally connected fireballs are on the order of $10$ to $100$ fm$^3$ for observables dependent on the formation of the light quark condensates, but could be larger for other observables.

\end{abstract}

\maketitle

\section{Introduction}\label{Intro}

Experiments in the field of high energy heavy ion collisions using very sophisticated detectors have been performed for 25 years, starting with the Relativistic Heavy Ion Collider (RHIC) at Brookhaven National Laboratory (BNL) in 2001. Variants of these experiments at even higher energies began at the Large Hadron Collider (LHC) at CERN in 2010. The purpose of these experiments is to investigate the properties of nuclear matter at temperatures and energy densities comparable to those present in the universe about a microsecond after the big bang. In these conditions, quantum chromodynamics (QCD), which describes the strong interactions between quarks and gluons, produces a phase change known as deconfinement. At temperatures typical of the contemporary universe, quarks are only found confined in hadrons; isolated quarks have never been observed. In the deconfinement phase of QCD, due to asymptotic freedom, quarks and gluons gain the ability to move individually forming a quark gluon plasma (QGP).

A fireball of QGP is created in the wake of a high energy collision between heavy nuclei. The fireball begins to rapidly cool and expand as soon as it is created, quickly forming a gas of hadrons as the nuclear matter within returns to the QCD confinement phase. During this process, the fireball can be treated as a relativistic, dissipative fluid \cite{Heinz:2013th,Gale:2013da}. In examining the physical properties of these fluids, it is often useful to model and quantify the sizes of the causal connections within, specifically in the central rapidity region. The study of these causal connections is described by Castorina and Satz as ``the counterpart of the cosmological horizon problem''; uniformity in the temperature of universal background radiation implies causal connections between regions of space thought to be causally disconnected \cite{CS}. Cosmological inflation theory was introduced in part to account for this discrepancy.  

The motivations for studying causal connections within QGP fireballs are threefold. First, if the causally connected volumes within a fireball in the central rapidity region are small enough at the beginning of hadron formation, globally conserved quantum numbers must also be locally conserved within many small causally connected regions, leading to the suppression of strange hadron production \cite{HamiehRedlichTounsi2000,KrausEtAl2007,KrausEtAl2009}. Second, the disoriented isospin condensate (DIC) mechanism, proposed as an explanation for anomalous neutral to charged kaon fluctuations observed by the ALICE collaboration at the LHC, depends on the causally connected volumes within a hadronizing QGP fireball being large enough for kaons produced in the central rapidity region to be correlated with one another \cite{ALICE:2021fpb,Kapusta:DIC,ChabowskiKapustaSingh2025}. Third, the chiral magnetic effect (CME) depends on the formation of parity-odd domains \cite{Dima1,Dima2,Dima3}; this effect may have been measured recently \cite{StarCME1,StarCME2}.

Castorina and Satz calculated the maximum causal connection of the fireball created in the wake of a high energy nucleus-nucleus collision in 1+1 dimensions using a mixture of Cartesian and Milne coordinates \cite{CS}. Milne coordinates are used to parameterize the Bjorken hydrodynamic flow model of QGP fireball evolution. The model contains one space and one time dimension. The spatial dimension described by Bjorken flow is along the direction of the particle collision beamline and is referred to as the longitudinal direction. The model assumes boost invariance in the longitudinal direction \cite{BJ}. Castorina and Satz found that a fireball produced with edges at proper time $\tau_1$ would, when its edges were at proper time $\tau_2$, be causally connected over a longitudinal distance $d = 2 z_2$, where $\tau_i^2 = t_i^2 - z_i^2$ and $z_2$ is given by the formula $d = \sqrt{\tau_2 / \tau_1} \left( \tau_2 - \tau_1 \right)$.

Our goal is to extend the work of Castorina and Satz by incorporating the transverse spatial directions into our description of the causal connections within QGP fireballs, thereby modeling the sizes of causal volumes in three spatial dimensions. In 2010, Gubser extended the Bjorken flow model into 3+1 dimensions by breaking translational symmetry in the transverse plane while preserving rotational symmetry. He derived the temperature function for this model using these symmetry constraints as well as the equation of state $P = \epsilon / 3$, corresponding to an ideal relativistic fluid \cite{Gubser1}. Gubser flow is the model we use in our investigation of causal connections in QGP fireballs.

We review Gubser flow as a modification of Bjorken flow in section \ref{GubserM}. We derive the trajectory of a Gubser fluid element in section \ref{MarkerP}. As a bonus, we obtain the proper time of every fluid element which, to our knowledge, has not appeared in the literature before. We calculate the transverse radius and overall volume of a causally connected fireball in section \ref{MaximumC}. Our formulas apply to central collisions between equal mass nuclei at energies high enough for boost invariance to be applicable in the central rapidity region.  The formulas can readily be extended to fireballs off the collision axis.  For concreteness, numerical results for Pb-Pb collisions at $\sqrt{s_{NN}} = 2.76$ TeV are shown.  The results are summarized and discussed in section \ref{Summa}.

\section{Gubser extension of Bjorken flow}\label{GubserM}

Bjorken flow is a boost-invariant hydrodynamic model that describes the behavior of the fluid created in the wake of a high energy heavy ion collision \cite{BJ}. It is conveniently expressed in Milne coordinates, which consist of proper time $\tau$ and spacetime rapidity $\eta$. In terms of Cartesian coordinates $\xi^\alpha = (t,x,y,z)$ they are
\ba
\tau &=& \sqrt{t^2 - z^2} \nonumber \\
\eta &=& \thalf \ln\left(\frac{t+z}{t-z}\right)
\ea
with the inverse relations
\ba
t &=& \tau \cosh\eta \nonumber \\
z &=& \tau \sinh\eta
\ea
The metric in Cartesian coordinates is
\be
\eta_{\alpha\beta}\, d\xi^\alpha d\xi^\beta = dt^2 -dx^2 -dy^2 - dz^2
\label{metricCart}
\ee
and in Milne coordinates
\be
g_{\mu\nu}\, dx^\mu dx^\nu = d\tau^2 - dx^2 -dy^2 - \tau^2 d\eta^2
\label{metricBJ}
\ee
The fluid element 4--velocity is 
\be
u^\mu = (\cosh\eta, 0, 0, \sinh\eta)
\ee
so that the speed of a fluid element is $\beta = \tanh\eta$.

Gubser generalized the boost invariant Bjorken hydrodynamics to include azimuthally symmetric transverse expansion but limited to the equation of state $P = \oneth \epsilon$ \cite{Gubser1,Gubser2}. In the transverse plane $x = r \cos\phi$ and $y = r \sin\phi$. The Milne coordinates $\tau$ and $\eta$ are defined the same as before. The solution is independent of $\phi$. The position in Cartesian coordinates, albeit expressed in terms of Milne variables, is
\be
x^\mu = \left( t, \, x, \, y, \, z \right) 
=\left( \tau \cosh\eta, \, r \cos\phi, \, r \sin\phi, \, \tau \sinh\eta \right) 
\label{4vec-x}
\ee
The 4--velocity of a fluid element, which we derive in Appendix \ref{AppA}, is
\be
u^\mu =
\left( \cosh\theta_\perp \cosh\eta, \, \sinh\theta_\perp \cos\phi, \, \sinh\theta_\perp \sin\phi, \, \cosh\theta_\perp \sinh\eta \right)
\label{eq:4vec-u}
\ee
where $\theta_\perp$ is a transverse rapidity.  It is given by
\be
\tanh \theta_\perp = \frac{2 q^2 \tau r}{1 + q^2 (\tau^2 + r^2)}\label{eq:TanhPerp}
\ee
The parameter $q$ is associated with the size in the transverse direction.  For central collisions of Au or Pb nuclei it has been estimated to be about $\oneqt$ fm$^{-1}$. The 3+1 dimensional Gubser hydrodynamics transforms back into 1+1 dimensional Bjorken hydrodynamics in the limit $q \rightarrow 0$. 

In the Gubser model, the radial speed is
\be
v_r = \frac{\tanh \theta_\perp}{\cosh\eta}\label{eq:TransVel}
\ee
Related hyperbolic functions are
\ba
\sinh \theta_\perp &=& \frac{2 q^2 \tau r}{\sqrt{1 + 2 q^2 (\tau^2 + r^2) + q^4 (\tau^2 - r^2)^2}} \nonumber \\
\cosh \theta_\perp &=& \frac{1 + q^2 (\tau^2 + r^2)}{\sqrt{1 + 2 q^2 (\tau^2 + r^2) + q^4 (\tau^2 - r^2)^2}}
\ea

Neglecting chemical potentials, the pressure is $P = a T^4$ where $a$ is a constant. The hydrodynamic equations of motion $\partial_\nu T^{\mu\nu} = 0$ can be written as \cite{J-F}
\ba
u^\mu \partial_\mu \ln T &=& - c_s^2 \partial_\mu u^\mu \label{eq:J-F1} \\
u^\nu \partial_\nu u^\mu &=& \left( g^{\mu\nu} - u^\mu u^\nu \right) \partial_\nu \ln T \label{eq:J-F2}
\ea
The solution to these hydrodynamic equations is \cite{J-F}
\be
T(\tau,r) = T_0 \left(\frac{\tau_0}{\tau}\right)^{1/3}
\frac{\left(1+q^2 \tau_0^2\right)^{2/3}}{\left[1 + 2 q^2 (\tau^2 + r^2) + q^4 (\tau^2 - r^2)^2\right]^{1/3}} 
\label{eq:J-F3}
\ee
Here $T(\tau_0,0) = T_0$, which is the initial temperature at proper time $\tau_0$ along the collision axis. It falls faster than the Bjorken result of $\left( \tau_0/\tau\right)^{1/3}$ due to transverse expansion. We provide an independent derivation of Eq. (\ref{eq:J-F3}) satisfying Eqs. (\ref{eq:J-F1}) and (\ref{eq:J-F2}) using Milne coordinates in Appendix \ref{AppA}.

\begin{figure}[H]
    \centering
    \includegraphics[width=0.7\linewidth]{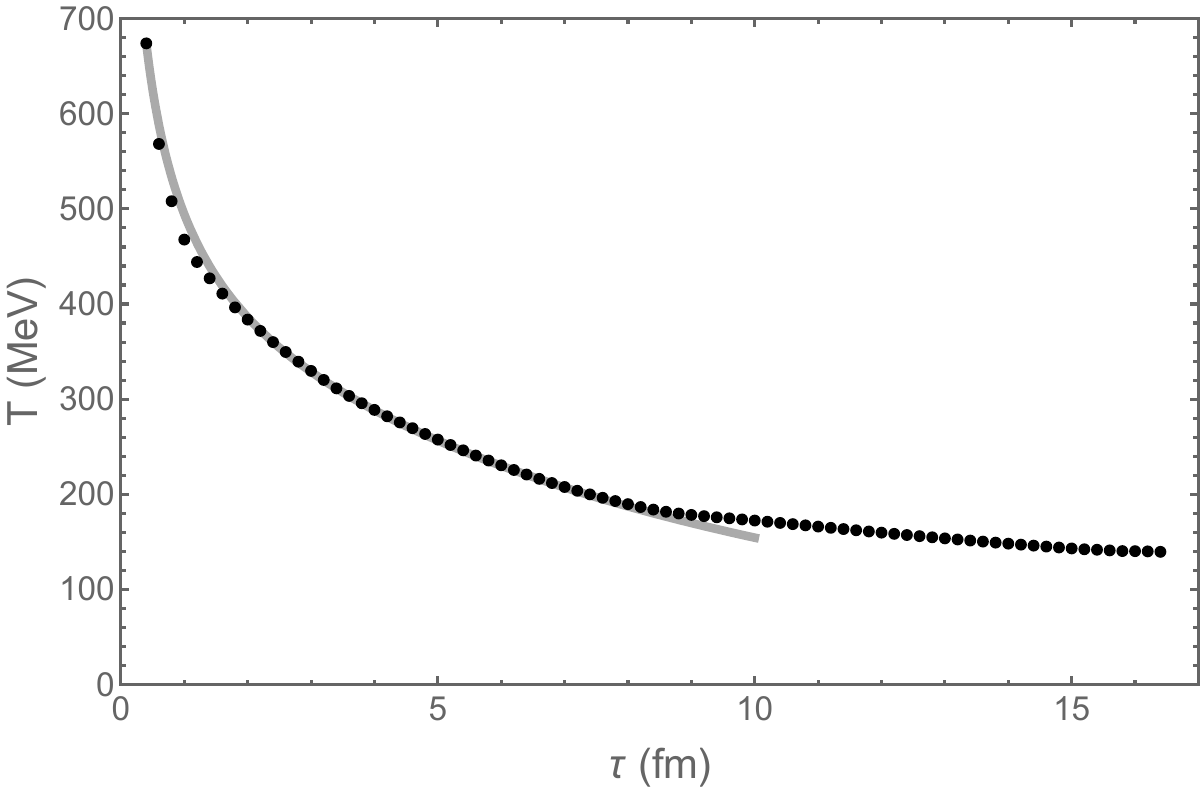}
    \caption{Simulated temperature evolution results data are shown as black points \cite{ChabowskiKapustaSingh2025}. The value of $T$ at $r = 0$ for a given value of $\tau$ according to Eq. (\ref{eq:J-F3}) with $\tau_0 = 0.4$ fm, $T_0 = 674$ MeV, and $q = 1/11$ fm$^{-1}$ is shown by the gray curve.}
    \label{fig:Ttpc}
\end{figure}

For the purposes of visualizing equations and performing calculations related to Gubser flow, we fix $\tau_0 = 0.4$ fm and $T_0 = 674$ MeV. These values were drawn from simulated temperature evolution results from numerical simulations of Pb-Pb collisions at $\sqrt{s_{NN}} = 2.76$ TeV using the IP-Glasma initial state \cite{Schenke:2012wb,McDonald:2016vlt} and hydrodynamic solver MUSIC \cite{Schenke:2010nt}.
In order to find a reasonable value for the parameter $q$, we used Eq. (\ref{eq:J-F3}) to plot T as a function of $\tau$ at $r = 0$ against the aforementioned simulated temperature evolution results. We then adjusted $q$ to best match the temperature as a function of time with the simulated temperature data. We focused on finding a strong match above $200$ MeV, which is where the equation of state used by Gubser is most likely to be valid. Based on Fig. \ref{fig:Ttpc}, we estimate that $1/11$ fm$^{-1}$ is an appropriate value for the parameter $q$.

\section{Gubser flow fluid element trajectory}\label{MarkerP}

Consider the transverse plane of the fireball at $\eta = z = 0$.  We want to find the trajectory $r(t)$ of a fluid element. From Eqs. (\ref{eq:TanhPerp}) and (\ref{eq:TransVel}), the relevant differential equation is
\be
\frac{dr}{d\tau} = v_r = \tanh\theta_\perp = \frac{2 q^2 \tau r}{1 + q^2 (\tau^2 + r^2)}
\ee
since $t = \tau$.  Multiply by $r/\tau$ to get
\be
\frac{dr^2}{d\tau^2} = \frac{2 q^2 r^2}{1 + q^2 (\tau^2 + r^2)}
\ee
Define $w = q^2 r^2$ and $v = 1 + q^2 \tau^2$.  The differential equation to solve is now
\be
\frac{dw}{dv} = \frac{2 w}{v + w}
\ee
This is a homogeneous differential equation of degree 1.  It can be put into separable form by writing $w = u v$.
\be
\frac{du}{dv} = \frac{u}{v} \left( \frac{1-u}{1+u} \right)
\ee
The solution is
\be
\frac{v}{v_0} = \frac{u}{u_0} \frac{(u_0-1)^2}{(u-1)^2}
\ee
where $u_0$ and $v_0$ represent the initial condition.  After solving the quadratic equation for $u(v)$ we find
\be
w = 2 c_0^2 + v - 2 c_0 \sqrt{c_0^2 + v}
\ee
where
\be
c_0 = \frac{1 + q^2(\tau_0^2 - r_0^2)}{2qr_0}
\ee
Finally
\be
qr(\tau) = -c_0 + \sqrt{c_0^2 + 1 + q^2 \tau^2} \label{eq:r_tau}
\ee
It is straightforward to verify that this satisfies the initial differential equation and the initial condition.

\begin{figure}[H]
    \centering
    \includegraphics[width=0.7\linewidth]{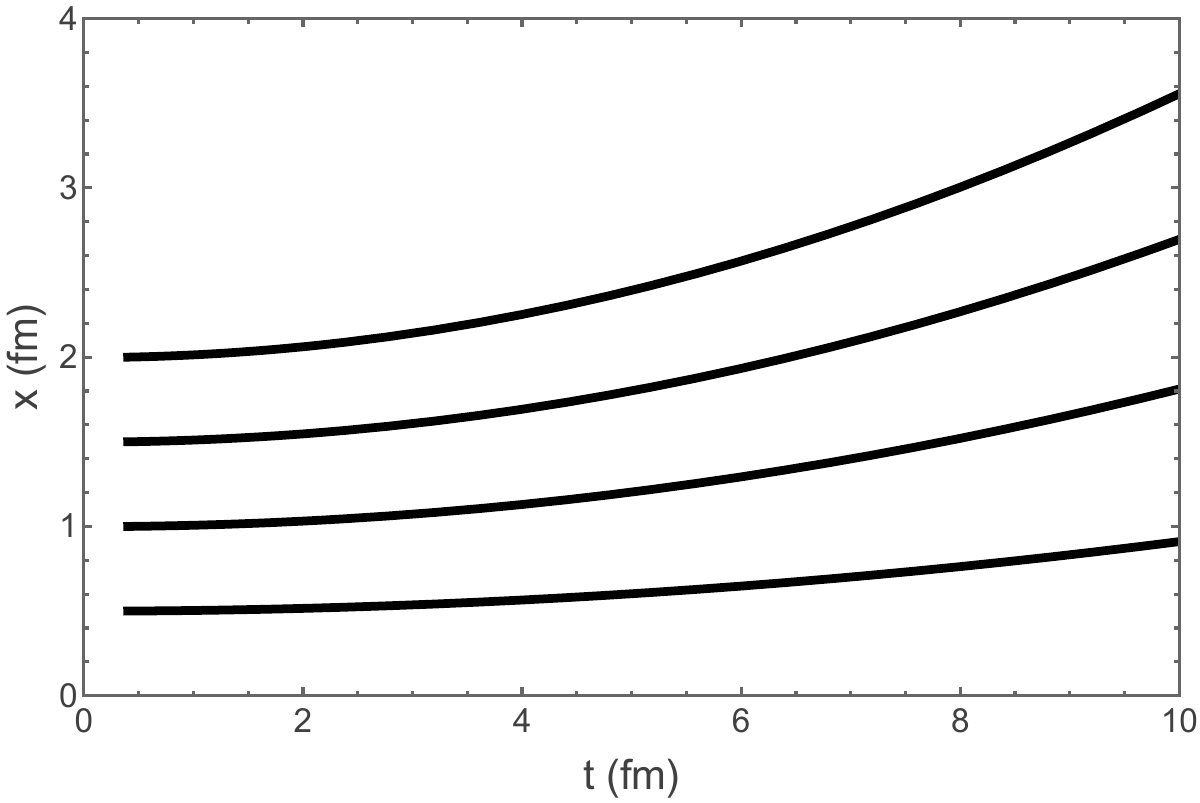}
    \caption{The path of a Gubser flow fluid element at $\eta = y = 0$. The input parameters are $\tau_0 = 0.4$ fm and $q = 1/11 \textrm{ fm}^{-1}$.}
    \label{fig:r_t_path}
\end{figure}

The relation between the proper time $\sigma$ of a fluid element and $\tau$ is determined by
\be
\frac{d \sigma}{d\tau} = \frac{1}{\cosh\theta_\perp(\tau,r(\tau))}
\ee
A straightforward substitution gives
\be
\frac{d \sigma}{d\tau} = \frac{\sqrt{c_0^2 + 1}}{\sqrt{c_0^2 + 1 + q^2 \tau^2}}
\ee
Integration gives
\be
\sigma(\tau) = \frac{\sqrt{c_0^2 + 1}}{q} \sinh^{-1}{\left(\frac{q \tau}{\sqrt{c_0^2 + 1}}\right)} + \sigma_0
\label{eq:sigma_tau}
\ee
and
\be
\tau(\sigma) = \frac{\sqrt{c_0^2 + 1}}{q} \sinh{\left( \frac{q (\sigma - \sigma_0)}{\sqrt{c_0^2 + 1}} \right)}
\ee
where $\sigma_0$ is a constant of integration.  As $r_0 \rightarrow 0$, $c_0 \rightarrow \infty$ and therefore
\be
\tau \rightarrow \sigma - \sigma_0
\ee
Since $\tau$ is the proper time for marker particles on the $z$ axis, $\sigma_0 = 0$. 

Consider what happens when $\eta \neq 0$.  A boost along the $z$ axis should not change the transverse spatial components.  Hence
\be
\sigma(\tau) =\frac{\sqrt{c_0^2 + 1}}{q} \sinh^{-1}{\left(\frac{q \tau}{\sqrt{c_0^2 + 1}}\right)}
\ee
Specifically, the components of the boost $\Lambda^\alpha_{\;\;\beta}$ are
\ba
\Lambda^0_{\;\;0} &=& \cosh\eta \nonumber \\
\Lambda^i_{\;\,0} &=& v_i \cosh\eta \nonumber \\
\Lambda^0_{\;\,j} &=& v_j \cosh\eta \nonumber \\
\Lambda^i_{\;j} &=& \delta_{ij} + (\cosh\eta - 1) \frac{v_i v_j}{v^2}
\ea
where $v_i = (0,0,\tanh\eta)$.  Boosting
\be
x^\mu_{\eta = 0} = \left( \tau, \, r(\tau) \cos\phi, \, r(\tau) \sin\phi, \, 0 \right) 
\ee
leads to
\be
x^\mu_\eta = \left( \tau \cosh\eta, \, r(\tau) \cos\phi, \, r(\tau) \sin\phi, \, \tau \sinh\eta \right) 
\ee
as expected.

\section{Maximum causal connections in Gubser flow}\label{MaximumC}

Castorina and Satz calculated the spatial extent of causally connected fireball resulting from a nucleus-nucleus collision in the context of Bjorken flow \cite{CS}. They considered a fireball created at $\tau_1$ which extends from $z_1$ to $-z_1$, where $z_1 < 0$. At a later time $\tau_2$ its ends have traveled out to greater distances $\pm z_2$, where $z_2 > 0$. The fireball will be a causally connected region of space if a light signal emitted at $z_1$ will just reach $z_2$. Castorina and Satz found that this will happen if the ends of the fireball are traveling with speeds $\pm \beta_{cc}$ where
\be
\beta_{cc} = \tanh\eta_{cc} = \frac{\tau_2 - \tau_1}{\tau_2 + \tau_1}\label{eq:beta_cc}
\ee
The $z_2$ value corresponding to Eq. (\ref{eq:beta_cc}) is
\be
z_2 = \tau_2 \sinh\eta_{cc} = \frac{\beta_{cc} \tau_2}{\sqrt{1 - \beta_{cc}^2}} = \frac{1}{2} \sqrt{\frac{\tau_2}{\tau_1}}(\tau_2 - \tau_1) \label{eq:z2}
 \ee
which is independent of the initial size $z_1$.  The total longitudinal extent of the fireball is thus $d = 2z_2$. Though this result was obtained in a system governed by Bjorken flow, it can be implemented in a system governed by Gubser flow as well due to the longitudinal boost invariance of both models. 

In the context of Gubser flow, the transverse expansion of a fireball must be taken into account. Consider a cross section in the transverse $x - y$ plane. Due to rotational symmetry we can set $y = 0$ for the present purpose. The fireball is centered at $\eta = 0$, so we set $\eta = 0$ such that $\tau = t$.

Point 1 is located at $x_1 < 0$ where the temperature has fallen to $T_1$ at some time $t_1$ determined by Eq. (\ref{eq:J-F3}). A light signal is sent to point 2 on the other side of the fireball which has temperature $T_2 < T_1$.  This point has coordinates $x_2 > 0$ and $t_2$ which are related by Eq. (\ref{eq:J-F3}). The trajectory of the light signal is
\be
x - x_1 = t - t_1
\ee
The condition that the light signal intersects point 2 is
\be
x_2 - x_1 = t_2 - t_1 \label{eq:inter}
\ee
There are now two conditions to determine $x_2$ and $t_2$, Eqs. (\ref{eq:J-F3}) and (\ref{eq:inter}). This will give the maximum causal connection.

\begin{figure}[H]
    \centering
    \includegraphics[width=0.7\linewidth]{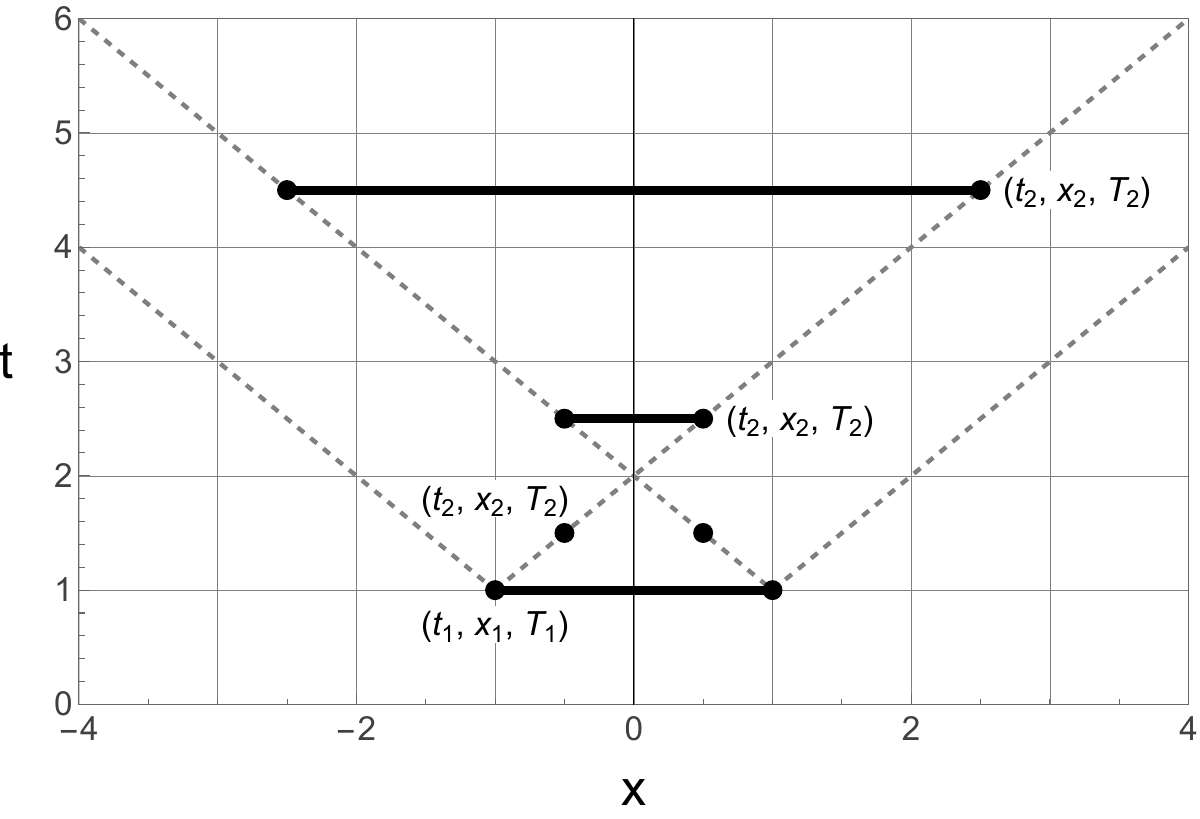}
    \caption{ The points $\left( t_1 , x_1 \right)$ and $\left( t_1 ,-x_1 \right)$ are at temperature $T_1$. The dashed lines represent the future light cones of these points. Three distinct possibilities are shown corresponding to three distinct values of $t_2$. The case $t_2 > t_1 - 2 x_1$ is displayed at $t_2 = 4.5$, wherein $x_2 > |x_1|$. In this case, the length of the region $-x_2 \leq x \leq x_2$ at $t_2$ is larger than that of the region $x_1 \leq x \leq |x_1|$ at $t_1$ and each point in one of these regions is causally connected to every point in the other. The case $t_1 - x_1 < t_2 < t_1 - 2 x_1$ is displayed at $t_2 = 2.5$, wherein $0 < x_2 < |x_1|$. In this case, the length of the region $-x_2 \leq x \leq x_2$ at $t_2$ is smaller than that of the region $x_1 \leq x \leq |x_1|$ at $t_1$ and each point in one of these regions is causally connected to every point in the other. The case $t_2 < t_1 - x_1$ is displayed at $t_2 = 2.5$, wherein $x_2 < 0$. In this case, there is no point at $t_2$, and therefore no point at temperature $T_2$, that is causally connected to every point in the region $x_1 \leq x \leq |x_1|$ at $t_1$. In all three of these cases, the points $\left( t_2 , -x_2 \right)$ and $\left( t_2 , x_2 \right)$ are at temperature $T_2$.}
    \label{fig:x2_cases}
\end{figure}

Having obtained a value of $t_1$ corresponding to a given pair of $T_1$ and $x_1$ values using Eq. (\ref{eq:J-F3}), we can substitute $t_2 = t_1 + x_2 - x_1$ into Eq. (\ref{eq:J-F3}) to obtain
\begin{align}
    \nonumber T_2 = \hspace{5pt} &T_0 \left(\frac{\tau_0}{t_1 + x_2 - x_1}\right)^{1/3}\\
    &\times \frac{\left(1+q^2 \tau_0^2\right)^{2/3}}{\left[1 + 2 q^2 ((t_1 + x_2 - x_1)^2 + x_2^2) + q^4 ((t_1 + x_2 - x_1)^2 - x_2^2)^2\right]^{1/3}} 
    \label{eq:J-F3-mod}
\end{align}
A numerical calculation can be done to obtain an $x_2$ corresponding to a given $T_2$.  If $x_2 > 0$, the transverse radius of the corresponding causally connected fireball is $r = x_2$. If $x_2 < 0$, no point on the fireball at $T_2$ is causally connected to both the point ($T_1$, $x_1$) and the point ($T_1$, $-x_1$).  Three possibilies are shown in Fig. \ref{fig:x2_cases}.  In this figure the $x_i$ and $t_i$ are unitless for the purpose of illustration.

Note that while Castorina and Satz imposed the condition in the longitudinal direction that the outer edges of the fireball needed to have equal velocities at times $t_1$ and $t_2$, we impose no equivalent condition in the transverse plane. Such a condition was necessary in the longitudinal direction to preserve longitudinal boost invariance. Transverse boost invariance is not a feature of Gubser flow.

\begin{figure}[H]
    \centering
    \includegraphics[width=0.7\linewidth]{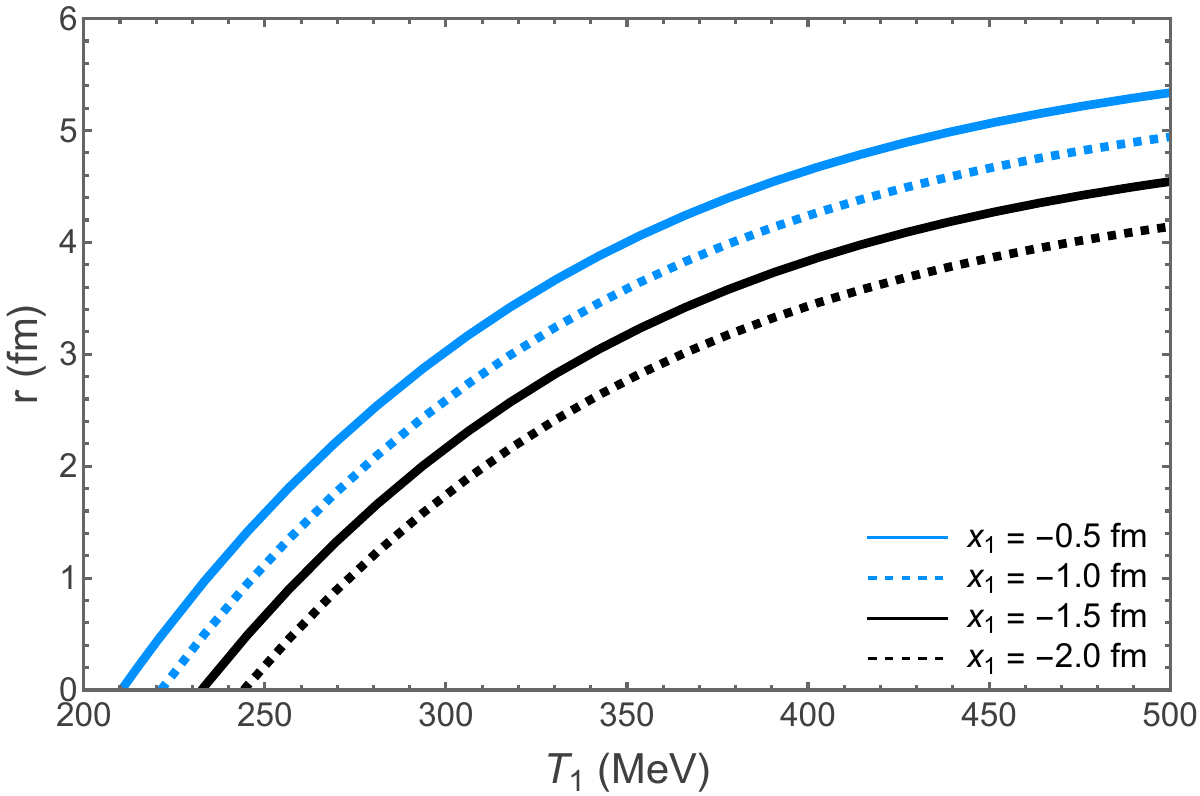}
    \caption{The transverse radius of a causally connected fireball in the $\eta = 0$ plane at $t_2$ with edges at $T_2 = 200$ MeV. The fireball is rotationally symmetric in the $x$-$y$ plane and had edges at $\pm x_1$ corresponding to $t_1$ and $T_1$. 
    }
    \label{fig:T2F200_11}
\end{figure}

\begin{figure}[H]
    \centering
    \includegraphics[width=0.7\linewidth]{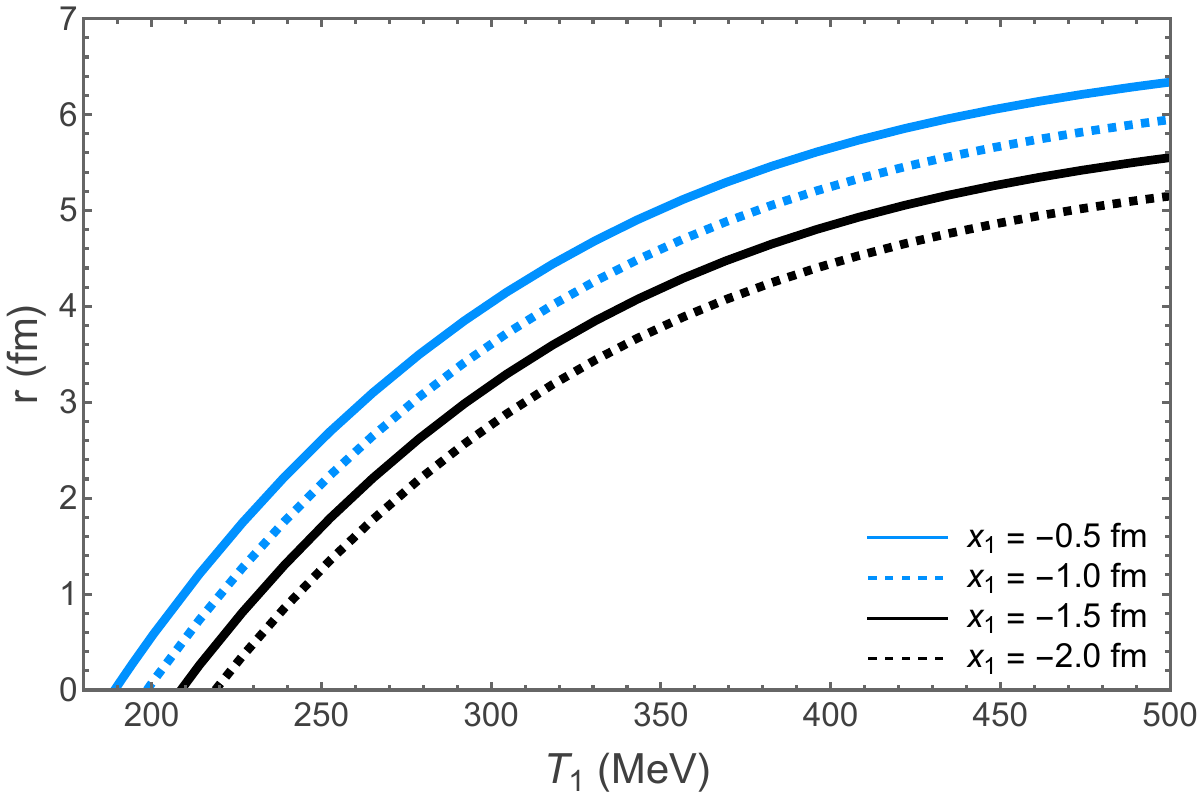}
    \caption{The transverse radius of a causally connected fireball in the $\eta = 0$ plane at $t_2$ with edges at $T_2 = 180$ MeV. The fireball is rotationally symmetric in the $x$-$y$ plane and had edges at $\pm x_1$ corresponding to $t_1$ and $T_1$. 
    }
    \label{fig:T2F180_11}
\end{figure}

We plot the transverse radius $r$ of a causally connected fireball as a function of $T_1$ in Figs. \ref{fig:T2F200_11} and \ref{fig:T2F180_11}. We choose 180 and 200 MeV as $T_2$ values because they fall within the range of temperatures at which cooling nuclear matter finishes its transition from QGP into hadron gas.

We can now estimate the causal volume of a fireball resulting from a Pb-Pb collision. Consider a fireball with transverse edges at $\pm x_1$ corresponding to $t_1$ and $T_1$, as well as longitudinal edges at $\pm z_1$ corresponding to $t_1$. We can use Eqs. (\ref{eq:z2}), (\ref{eq:inter}), and (\ref{eq:J-F3-mod}) to find the transverse radius $r$, corresponding to $t_2$ and $T_2$, and the longitudinal extent $d = 2 z_2$, corresponding to $t_2$, of the causally connected region of the fireball. To obtain the appropriate value of $z_2$, we transform Castorina and Satz's result, Eq. (\ref{eq:z2}), into Cartesian coordinates, thereby obtaining
\begin{equation}
    z_2 = \frac{t_2 \left( t_2 - t_1 \right)}{t_1 + t_2}
    \label{eq:z2finder}
\end{equation}
As expected, the value of $z_2$ does not depend explicitly on $z_1$. We use Eq. (\ref{eq:z2finder}) in concert with Eq. (\ref{eq:J-F3-mod}) to find $r = x_2$ and $d = 2 z_2$ for a given set of $T_1$, $T_2$, and $x_1$ values.

\begin{figure}[H]
    \centering
    \includegraphics[width=0.7\linewidth]{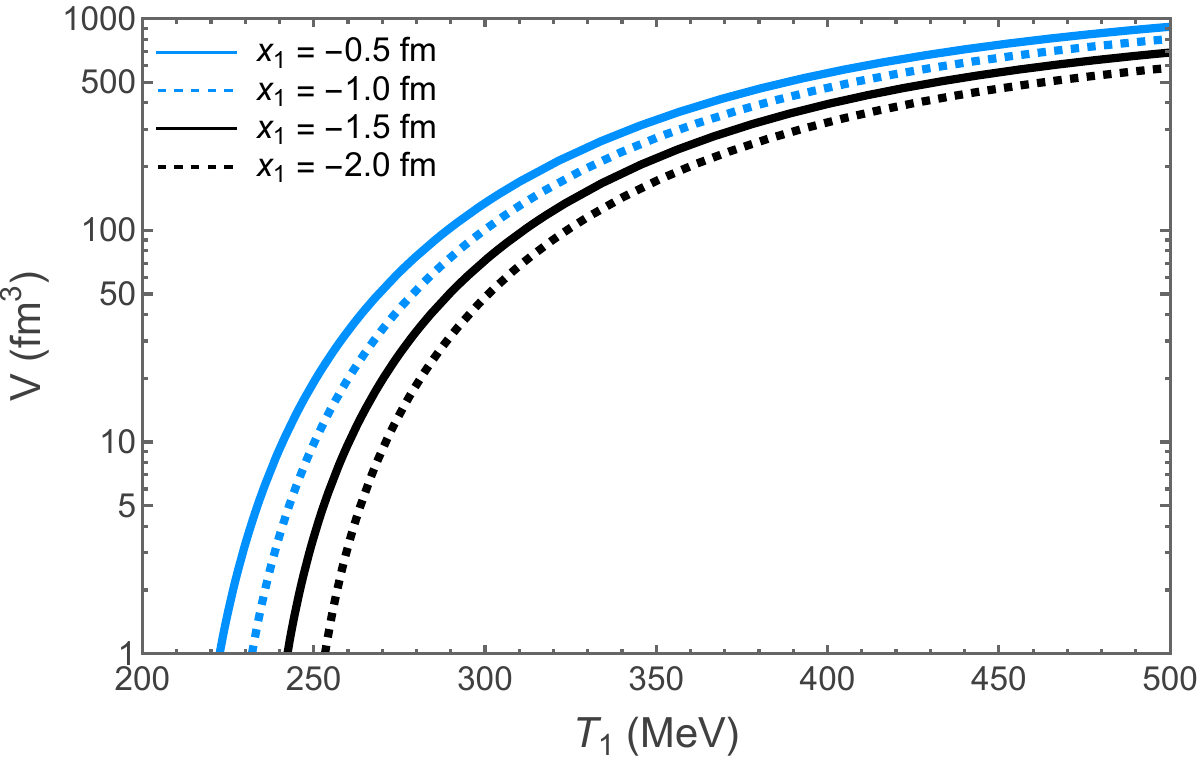}
    \caption{The spatial volume of a causally connected fireball with transverse edges at $T_2 = 200$ MeV corresponding to time $t_2$ and longitudinal extent $d = 2 z_2$. The fireball had transverse edges at $\pm x_1$ corresponding to $t_1$ and $T_1$. 
    }
    \label{fig:vol500}
\end{figure}

\begin{figure}[H]
    \centering
    \includegraphics[width=0.7\linewidth]{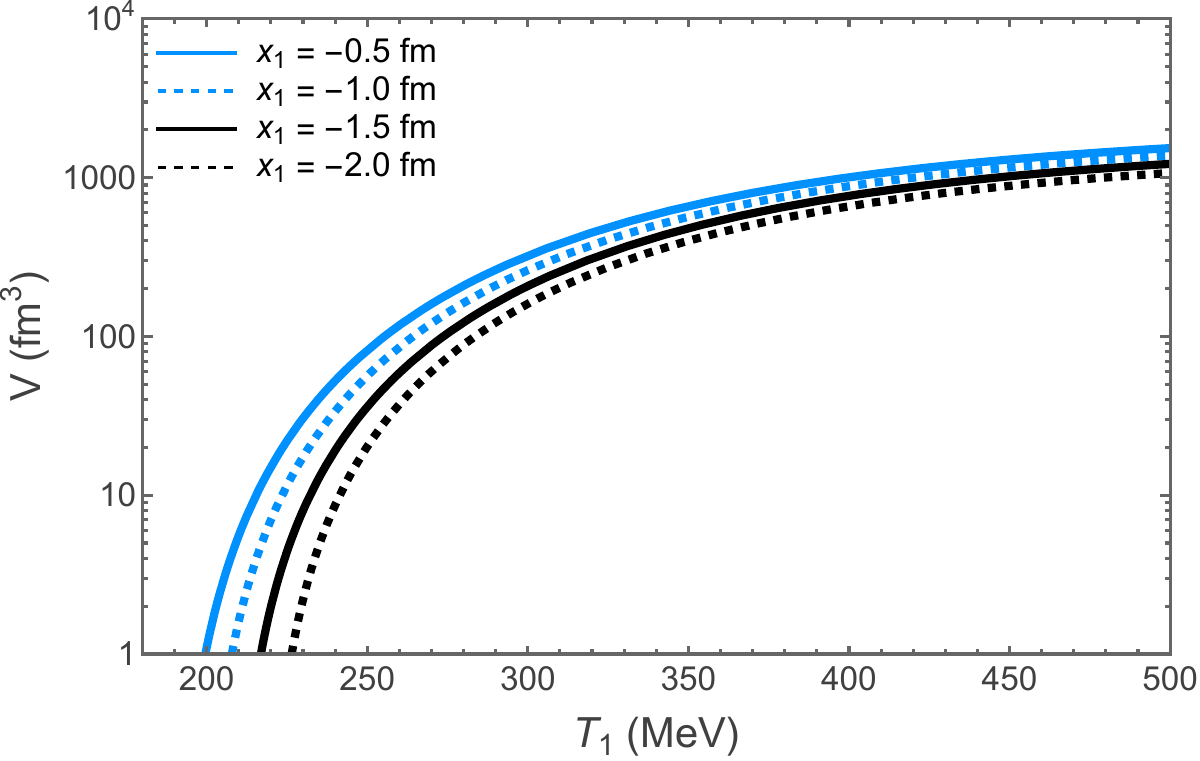}
    \caption{The spatial volume of a causally connected fireball with transverse edges at $T_2 = 180$ MeV corresponding to time $t_2$ and longitudinal extent $d = 2 z_2$. The fireball had transverse edges at $\pm x_1$ corresponding to $t_1$ and $T_1$.
    }
    \label{fig:180vol500}
\end{figure}

Approximating the shape of this causally connected fireball in spatial coordinates to be a cylinder of length $d = 2 z_2$ and radius $r$ at time $t_2$, we obtain $V= \pi r^2 d $, the spatial volume of the fireball. Fireball volumes are shown in Figs. \ref{fig:vol500} and \ref{fig:180vol500}. These figures suggest that the causal volumes of these fireballs are on the order of $10$ to $100$ fm$^3$ when $T_1 \approx 275$ MeV. We mention this temperature because that is when the light quark condensate could conceivably begin forming, which is relevant for the DIC \cite{Kapusta:DIC}. The specific temperature at which the condensate begins forming is unclear as this phase change is a smooth crossover transition.

\section{Summary and conclusions}\label{Summa}

We extended the analysis of Castorina and Satz's calculation of the volume of a causally connected region in a central high energy nucleus-nucleus collision using the Gubser extension of the Bjorken model, which includes transverse expansion. Unlike the longitudinal calculation performed by Castorina and Satz, the transverse calculation depends on the transverse radius at the initial time of interest. We gave three examples of observables which depend on this volume and gave numerical examples for one of them (DIC).  As a bonus, we found the proper time of any fluid element in terms of the Milne proper time and the radius.  This analysis can easily be extended to regions that are not centered on the beam axis.  Very challenging will be extending this analysis to arbitrary 3+1 dimensional simulations at nonzero impact parameters, such as in Refs. \cite{Heinz:2013th,Gale:2013da}. 

\begin{acknowledgments}
We thank J.-F. Paquet, C. Plumberg, and H. Satz for advice and feedback on this manuscript. This work was supported by the U.S. DOE Grant No. DE-FG02-87ER40328.  It was prepared without the use of large language models (LLMs) or any other forms of generative artificial intelligence (GenAI).
\end{acknowledgments}

\appendix

\section{Gubser flow derivations}\label{AppA}

In order to derive the fluid element 4--velocity in Gubser flow, we utilize the 4--velocity profile that respects the symmetries of Gubser flow, namely \cite{Gubser1}
\begin{equation}
    u = \cosh \theta_\perp \frac{\partial}{\partial \tau} + \sinh \theta_\perp \frac{\partial}{\partial r}
\end{equation}
If we take $t$ and $z$ to be defined using the familiar conventions of Milne coordinates such that $t = \tau \cosh \eta$ and $z = \tau \sinh \eta$, then we obtain $u^\mu$ components that match Eq. (\ref{eq:4vec-u}).
\begin{align}
    \nonumber u^t &= \cosh \theta_\perp \frac{\partial}{\partial \tau}(\tau \cosh \eta) + \sinh \theta_\perp \frac{\partial}{\partial r}(\tau \cosh \eta) = \cosh \theta_\perp \cosh \eta\\
    \nonumber u^x &= \cosh \theta_\perp \frac{\partial}{\partial \tau}(r \cos \phi) + \sinh \theta_\perp \frac{\partial}{\partial r}(r \cos \phi) = \sinh \theta_\perp \cos \phi\\
    \nonumber u^y &= \cosh \theta_\perp \frac{\partial}{\partial \tau}(r \sin \phi) + \sinh \theta_\perp \frac{\partial}{\partial r}(r \sin \phi) = \sinh \theta_\perp \sin \phi\\
    u^z &= \cosh \theta_\perp \frac{\partial}{\partial \tau}(\tau \cosh \eta) + \sinh \theta_\perp \frac{\partial}{\partial r}(\tau \cosh \eta) = \cosh \theta_\perp \sinh \eta
\end{align}

In order to verify that Eq. (\ref{eq:J-F3}) is a solution to Eqs. (\ref{eq:J-F1}) and (\ref{eq:J-F2}), we begin by finding the relationships between partial derivatives in Cartesian and Milne coordinates, as well as by finding the co--moving time derivative and divergence of the Gubser flow 4--velocity. The metric is taken to be $g_{\mu\nu} = {\rm diag} (1, -1, -1, -1)$. In terms of Milne coordinates the $t$ and $z$ derivatives are
\ba
\frac{\partial}{\partial t} &=& \cosh\eta \frac{\partial}{\partial \tau} 
- \frac{\sinh\eta}{\tau} \frac{\partial}{\partial \eta} \\
\frac{\partial}{\partial z} &=& - \sinh\eta \frac{\partial}{\partial \tau} 
+ \frac{\cosh\eta}{\tau} \frac{\partial}{\partial \eta}
\ea
The co--moving time derivative is
\be
u^\mu \partial_\mu = \cosh\theta_\perp \frac{\partial}{\partial \tau} 
+ \sinh\theta_\perp \frac{\partial}{\partial r}
\ee
and the divergence of the 4--velocity is
\ba
\partial_\mu u^\mu &=& \frac{\partial}{\partial \tau} \cosh\theta_\perp + \frac{\cosh\theta_\perp}{\tau}
+ \frac{\partial}{\partial r} \sinh\theta_\perp + \frac{\sinh\theta_\perp}{r} \nonumber \\
&=& \frac{1}{\tau} \frac{\partial}{\partial \tau} \left( \tau \cosh\theta_\perp \right)
+ \frac{1}{r} \frac{\partial}{\partial r} \left( r \sinh\theta_\perp \right) \nonumber \\
&=& \frac{\cosh\theta_\perp}{\tau} + \frac{4 q^2 \tau}{\alpha}
\label{eq:div_4vel}
\ea
where $\alpha \equiv \sqrt{1 + 2 q^2 (\tau^2 + r^2) + q^4 (\tau^2 - r^2)^2}$.

From Eq. (\ref{eq:div_4vel}), the RHS of Eq. (\ref{eq:J-F1}) is equal to
\begin{equation}
    -c_s^2 \left( \frac{\cosh\theta_\perp}{\tau} + \frac{4 q^2 \tau}{\alpha} \right)
\end{equation}
Plugging Eq. (\ref{eq:J-F3}) into the LHS of Eq. (\ref{eq:J-F1}), we find
\begin{equation}
    u^\mu \partial_\mu \ln T = u^\mu \partial_\mu \left( -\ln \tau^{1/3} - \ln \alpha^{2/3} \right) = -\frac{1}{3} \left( \frac{\cosh\theta_\perp}{\tau} + \frac{4 q^2 \tau}{\alpha} \right)
\end{equation}
Therefore Eq. (\ref{eq:J-F1}) is satisfied by Eq. (\ref{eq:J-F3}) if $c_s^2 = \oneth$, which is the conformal equation of state assumed in deriving the Gubser flow solution.

Plugging Eq. (\ref{eq:J-F3}) into the $t$-component of Eq. (\ref{eq:J-F2}), we find 
\begin{align}
    \nonumber &\left(g^{t \nu} - u^t u^\nu \right) \partial_\nu \ln T = g^{t t} \partial_t \ln{T} - u^t u^\nu \partial_\nu \ln{T}\\ 
    \nonumber &= \cosh\eta \frac{\partial \ln{T}}{\partial \tau} + \frac{1}{3} \left( \cosh\theta_\perp \cosh{\eta} \right) \left( \frac{\cosh\theta_\perp}{\tau} + \frac{4 q^2 \tau}{\alpha} \right) = \frac{ \sinh^2{\theta_\perp} \cosh{\eta}}{\tau}\\ 
    &= \cosh\theta_\perp \cosh{\eta} \frac{\partial}{\partial \tau}\left( \cosh\theta_\perp \right) + \sinh\theta_\perp \cosh{\eta} \frac{\partial}{\partial r}\left( \cosh\theta_\perp \right) = u^\nu \partial_\nu u^t
\end{align}
For the $x$-component of Eq. (\ref{eq:J-F2}), we find
\begin{align}
    \nonumber &\left( g^{x \nu} - u^x u^\nu \right) \partial_\nu \ln T = g^{x x} \partial_x \ln{T} - u^x u^\nu \partial_\nu \ln{T}\\ 
    \nonumber &= - \cos\phi \frac{\partial \ln{T}}{\partial r} + \frac{1}{3} \left( \cos\phi \sinh\theta_\perp \right) \left( \frac{\cosh\theta_\perp}{\tau} + \frac{4 q^2 \tau}{\alpha} \right) = \frac{\cos{\phi} \cosh\theta_\perp \sinh\theta_\perp}{\tau}\\
    &= \cosh\theta_\perp \cos{\phi} \frac{\partial}{\partial \tau}\left( \sinh\theta_\perp \right) + \sinh\theta_\perp \cos{\phi} \frac{\partial}{\partial r}\left( \sinh\theta_\perp \right) = u^\nu \partial_\nu u^x
\end{align}
The $y$-component of Eq. (\ref{eq:J-F2}) is similar to the $x$-component.
\begin{equation}
    \left( g^{y \nu} - u^y u^\nu \right) \partial_\nu \ln T = \frac{\sin{\phi} \cosh\theta_\perp \sinh\theta_\perp}{\tau} = u^\nu \partial_\nu u^y
\end{equation}
Finally, for the $z$-component of Eq. (\ref{eq:J-F2}), we find
\begin{align}
    \nonumber &\left( g^{z \nu} - u^z u^\nu \right) \partial_\nu \ln T = g^{z z} \partial_z \ln{T} - u^z u^\nu \partial_\nu \ln{T}\\ 
    \nonumber &= \sinh\eta \frac{\partial \ln{T}}{\partial \tau} + \frac{1}{3} \left( \cosh\theta_\perp \sinh{\eta} \right) \left( \frac{\cosh\theta_\perp}{\tau} + \frac{4 q^2 \tau}{\alpha} \right) = \frac{\sinh^2{\theta_\perp} \sinh{\eta}}{\tau}\\
    &= \cosh\theta_\perp \sinh{\eta} \frac{\partial}{\partial \tau}\left( \cosh\theta_\perp \right) + \sinh\theta_\perp \sinh{\eta} \frac{\partial}{\partial r}\left( \cosh\theta_\perp \right) = u^\nu \partial_\nu u^z
\end{align}
Therefore, Eq. (\ref{eq:J-F3}) satisfies Eq. (\ref{eq:J-F2}).

\end{document}